# Rapture in the Cartesian Wall between Real World Entities and their Abstract Models

Narada Wickramage
IT & MIS Division
Public Utilities Commission of Sri Lanka
Colombo, Sri Lanka
naradaw@pucsl.gov.lk

*Abstract* — *This short paper envisages that the advancements made with respect to Big Data (BD), High Performance Computing, etc. would give rise to a new paradigm of concrete information models, which would closely replicate the real world and the consequences such as self-verifying information models, BD warehouses as intermediaries between data sources and information systems, etc.*

## I. INTRODUCTION

Ever since man understood the ability of using machines to process data, he has been able to make significant progress as far as creating abstract representations of real world or hypothetical scenarios inside computing machines. Abstraction is necessary because real world is so complex that replicating an entire real world scenario inside a machine is not efficient, economical or practical. Computer Engineers are ever burdened with building viable information models, which can sufficiently represent real world scenarios so that a finite amount of data can be processed within a finite amount of time to produce the expected output.

A consumer modeled within a customer relationship management system, a patient modeled within a health information system, a DNA molecule modeled within a bioinformatics system and a star modeled within a cosmological information system are all examples of abstract models which can be way different from their real world counterparts. What prevent them from becoming fuller and timely representations of the real world entities, which they are supposed to stand for, are the difficulties pertaining to gathering necessary information to construct more accurate information models and limitations in computing power to process such enormous volumes of data fast enough.

Due to these constraints, an information system pertaining to customer relationship management might use either an abstract model of a general customer, who in turn represents an entire market segment or abstract models of each and every member of that market segment. In either way, these models persistently lack all the information that can be found in the physical world, which is said to be simulated inside various information systems.

Quantitative changes taking place in the fields such as numerous sources of data gathering, generation & retrieval, Big Data (BD), High Speed Networks (HSN), High Performance Computing (HPC), terabyte memory chips, Exabyte storage, High Performance Data Analysis (HPDA), etc. can be expected to effect qualitative changes as far as the way natural world is modeled inside computing systems is concerned. With the narrowing of the gap between electronic simulations and real world processes as far as the amount of information present in virtual models is concerned, the digital representations of real world entities become true and complete reproductions of their actual counterparts.

With the rapture if not the fall of the Cartesian wall between the physical phenomena and virtual representations of them, the cleaner abstractions, which dominate contemporary computing paradigm, would be replaced with concrete reproductions of actual entities present in the natural world. While the relationship between complexity and simplicity is complex [1], the absence of the Cartesian wall resulting in due to a paradigm shift towards information rich electronic representations would invariably allow complexities inherent in the real world to invade the virtual world, which is currently consisting of simpler, cleaner and smoother generalizations of actual phenomena.

However this does not imply that the degree of abstractness of an information system, considered in its entirety, would decrease. On the contrary, the level of abstractness, which is at the heart of computing, would in fact increase due to the attempts to make predictions, etc. by processing rich data sets. But the information models that represent the real world phenomena would closely match what they stand for. In certain situations, this would increase the number of information models used in an information system due to the lack of generality of each information model.

The next section of the paper explains the factors that precipitate this qualitative development, whereas the third section discusses the repercussions of the paradigm shift. The final section describes areas for further exploration.

## II. CAUSATIVE FACTORS OF THE PARADIGM SHIFT

Sources of BD are growing. Data from a Vast range of fields such as data generated due to human participation in Social Networks (SN), data gathered from sensor networks, data collected from various SaaS services, data from emails, short messages, comments posted in blogs, etc. data from satellites and map applications, data generated out of multimedia files such as photos, videos, audios uploaded, data accumulated in data warehouses due to business transaction, etc. are piling up huge Volumes of BD. Socially enabled persons, whose "entire biography" since their birth is available online, the Internet of Things, which would generate an unimaginable amount of information, etc. would construct BD Warehouses of astronomical proportions.

One important implications of accumulating and datafying [2] information from multiple sources is BD become Verifiable so that Veracity and Validity of BD can be established and it is just a matter of time before mechanisms would evolve to cross check Variety of data emanating from Various sources against one another to assure consistency of BD.

Volant data transmitted quickly via HSN and the Velocity with which Volumes of them processed because of HPC facilities would ensure that not only information rich and almost-completed replicas (as opposed to mere skeletons) of real world entities containing current data (as opposed to obsolete data) would be built as components of electronic simulations but also the information contained in such replicas would be more accurate.

While the existence of data from plethora of sources induces data verification, the data verification process would cyclically affect the information in BD repositories so that reliability of BD would further be improved. This in turn further helps constructing information models which contain correct data retrieved from BD repositories. Currently SNs are trending towards ensuring the accuracy, authenticity and accessibility of data stored in them and this would be the norm for almost every source of information. This would further enhance the dependability of data in BD repositories.

III. REPERCUSSIONS

An important feature pertaining to the models in the new paradigm is the mechanisms associated with such models to gather information from numerous sources in the outside world and cross check them in order to verify them. To illustrate this with an example from Object Oriented Programming (OOP), in addition to mutators (setters) and accessors (getters) present in OOP, Classes would have gatherers and verifiers, which would gather additional information from BD repositories and verify them using semi-intelligent algorithms.

The clouds would host seas of BD which contain enormous volumes of data from gamut of sources and like SOAP, WSDL, UDDI, etc. in Web services, a range of specifications would describe interfaces and protocols that would facilitate feeding data to BD Warehouses from sources mentioned in the 2$^{nd}$ section and retrieving data from BD repositories by information models and interpreted in such a manner that last mile of data analysis is achieved [3].

A third implication would be the quasi-intelligent mechanisms to develop hypotheses that would be tested as a way to gauge the level of trustworthiness of data contained in information models so that the gap between the real world scenario and the corresponding simulation can be measured and narrowed. Based on the data contained in a model, testable suppositions can be derived and for instance, based on one's status updates mentioned in a certain social networking site, one's interests can be guessed and it is possible to find out whether data from e-commerce transactions corroborates the assumption made regarding the one's interests.

Hypothesis testing would estimate how concrete the model is and until the model is sufficiently concrete, data in the information model would be replaced with data from alternative sources, enriched with additional data and perfected through data processing.

Since it is reasonable to expect that data from all sources would not be equally sound, mechanisms would evolve to rank data sources based on merit and earmark either good or bad sources of data enumerated in a registry and facilitate discovery of sources, used to retrieve information from, based on rank, which is determined in a crowed sourced manner.

The concreteness of one information model within an information system would influence the concreteness of other models as part of the data verification process and this would in turn facilitate the confirmation procedure. The emphasis on the concreteness of information models in the new paradigm ensures that models would contain data more than what is necessary unlike the objects used in OOP and this would in turn create further repercussions.

When information models have surplus of data, models can engage in HPDA facilitated by HPC & HSN infrastructure (including attempts to substantiate hypotheses) and this can augment the behavior of concrete information models.

IV. EXPLORABLE AREAS

The shift from the present abstract-model paradigm to new concrete-model paradigm would generate plenty of research topics. How BD can be solidified to increase its consistency without losing diversity by improving on the soundness of data will be probed. Discovering suitable data repositories, retrieving data from them and enhancing information models would be an area of study. Defining interfaces of BD warehouses and protocols to interact with them would be explored. Developing algorithms to evolve hypotheses would be researched. Methods of interpreting BD would be a field of investigation. An area of research would be to understand how the behavior of data rich information models can be boosted by using extra information present in models. Developing intelligent and autonomous object models will be investigated.